\documentclass[12pt]{article}

\usepackage{amsmath}
\usepackage{amssymb}
\usepackage{geometry}
\usepackage{setspace}
\usepackage{hyperref}
\usepackage{graphicx}
\usepackage{booktabs}
\usepackage{natbib}

\title{Weighted $k$-Sample Kolmogorov–Smirnov, Cramér–von Mises, and Anderson–Darling Tests for Assessing Covariate Balance}

{\small
\author{Ariel Linden, DrPH\\
University of California, San Francisco\\
Department of Medicine\\
Division of Clinical Informatics \& Digital Transformation (DoC-IT)\\
San Francisco, CA, USA\\
ariel.linden@ucsf.edu}
}

\date{}
\begin{document}

\maketitle

\section*{Abstract}
Weighted distributional tests for covariate balance are currently limited to two-group comparisons. We extend the Kolmogorov--Smirnov, Anderson--Darling, and Cram\'er--von Mises tests to an arbitrary number of weighted groups $k \ge 2$, using existing $k$-sample generalizations and a shared permutation-inference procedure; each statistic reduces exactly to its two-group counterpart at $k=2$. An accompanying post-hoc pairwise procedure with four multiple-comparison adjustments localizes which groups differ following an omnibus rejection. In a four-scenario simulation study at $k=3$, Type~I error remained close to nominal, and the comparative advantages established for two groups were preserved: Kolmogorov--Smirnov was most powerful against a centrally located discrepancy, Anderson--Darling against a tail-located discrepancy, and Anderson--Darling and Cram\'er--von Mises performed comparably against a diffuse discrepancy. Omnibus power was nonetheless uniformly lower than in the matching two-group setting --- not because the underlying discrepancy is diluted, but because adding groups enlarges the null distribution itself, so the post-hoc procedure, whose pairwise statistics carry no such penalty, is often the more sensitive tool whenever a specific group's imbalance is suspected. The methods are implemented in the Stata commands \texttt{kstest}, \texttt{adtest}, and \texttt{cvmtest}.

\section*{Keywords}

covariate balance; distributional tests; $k$-sample tests; Kolmogorov--Smirnov test; Anderson--Darling test; Cram\'er--von Mises test; permutation inference; weighting; post-hoc comparisons

\section{Introduction}

Balance testing is an integral component of causal inferential methods. Covariate balance may be achieved by design, as in randomized experiments, or by adjustment, as in propensity-score weighting \citep{robins2000}, stratification \citep{linden2014}, entropy balancing \citep{Hainmueller2012}, or matching. Either way, demonstrating that comparison groups are comparable on observed, measured characteristics is a necessary step in supporting a causal interpretation of an estimated treatment effect \citep{rubin2008}. In a growing share of applied settings, this comparison is not between two groups but among three or more: multi-arm randomized trials, treatments with more than two nominal categories (e.g., dose levels, intervention modalities, or exposure intensities), and observational studies with multiple distinct comparison groups.

\citet{linden2026arxiv} recently extended three classical goodness-of-fit tests --- Kolmogorov--Smirnov (KS) \citep{kolmogorov1933,smirnov1948}, Anderson--Darling (AD) \citep{anderson1952}, and Cram\'er--von Mises (CVM) \citep{cramer1928,vonmises1931} --- to accommodate case weights of any origin, using a shared label-permutation inference procedure. That extension, like the classical tests themselves, was limited to two-group comparisons. When balance must be assessed across three or more groups, an investigator working from those methods is left with two unsatisfying options: run every pairwise comparison separately, with no principled way to control the resulting multiple-comparisons problem, or fall back to a summary measure (e.g., a standardized mean difference computed pairwise or against a reference group) that discards the distributional information the two-group extension was built to preserve \citep{linden2016multivalued}.

This paper extends the weighted two-group framework to an arbitrary number of groups $k \ge 2$. We use two existing $k$-sample generalizations from the classical (unweighted) goodness-of-fit literature: Kiefer's (1959) $k$-sample statistics for KS and CVM \citep{kiefer1959}, and the discrete $k$-sample Anderson--Darling statistic of Scholz and Stephens \citep{scholz1987}. Each is combined with case weights and the same weight-agnostic permutation-inference procedure used in the two-group extension, and each is constructed so that it reduces \emph{exactly} to its two-group counterpart when $k=2$ --- not merely asymptotically or up to a scaling constant, but as the identical statistic, computed the identical way. This means the two-group tests are not superseded by this extension; they are a special case of it.

A second contribution addresses a question that has no two-group analogue: once an omnibus $k$-sample test indicates that \emph{some} group's distribution differs from the others, which group, or which pair of groups, is responsible? We introduce a post-hoc pairwise comparison procedure built directly on the same permutation machinery, together with a choice of four standard multiple-comparison adjustments, so that an investigator can move from ``these $k$ groups are not all balanced'' to a specific, adjusted answer about which pairs differ, without leaving the same framework or introducing a different, unrelated method for that second question.

As in the two-group setting, the three tests are not interchangeable: they differ systematically in their sensitivity to \emph{where} a distributional discrepancy is located, a property that follows from each statistic's construction \citep{stephens1974,dagostino1986} and that was previously confirmed to survive weighting in the two-group case \citep{linden2026arxiv}. This paper asks whether that same pattern --- KS most sensitive to a centrally located discrepancy, AD most sensitive to a tail-located discrepancy, and AD/CVM comparably sensitive to a diffuse discrepancy --- persists once the comparison is extended from two groups to three, under the same kind of case-weighting and the same permutation-inference procedure. The methods are implemented in the freely available, community-contributed Stata commands \texttt{kstest} \citep{lindenkstest2026}, \texttt{adtest} \citep{lindenadtest2025}, and \texttt{cvmtest} \citep{lindencvmtest2025}.

\section{Methods}

\subsection{Notation and General Framework}

Consider $k \ge 2$ independent samples, denoted Groups $1,\ldots,k$, of sizes $n_1,\ldots,n_k$, drawn from continuous distributions with cumulative distribution functions $F_1,\ldots,F_k$. The null hypothesis of interest is the homogeneity hypothesis $H_1: F_1 = F_2 = \cdots = F_k$ for a given covariate: that is, that the covariate's distribution is balanced across all $k$ groups once weighting has been applied. When $k=2$, this reduces to the two-group null hypothesis $H_0: F_1=F_0$ studied by \citet{linden2026arxiv}.

Each observation $i$ in group $j$ carries a case weight $w_{ji} > 0$. The weighted empirical cumulative distribution function (ECDF) for group $j$ is
\begin{equation}
\hat{F}_j^{\,w}(x) \;=\; \frac{\sum_{i=1}^{n_j} w_{ji}\,\mathbb{1}(x_{ji} \le x)}{\sum_{i=1}^{n_j} w_{ji}},
\end{equation}
exactly as in the two-group case. We additionally define the weighted \emph{pooled} ECDF across all $k$ groups,
\begin{equation}
\bar{F}^{\,w}(x) \;=\; \frac{\sum_{j=1}^{k}\sum_{i=1}^{n_j} w_{ji}\,\mathbb{1}(x_{ji}\le x)}{\sum_{j=1}^{k}\sum_{i=1}^{n_j} w_{ji}},
\end{equation}
which plays the role in the $k$-sample statistics below that the second group's ECDF, $\hat F_0^w$, played directly in the two-group statistics: every group is compared not to one other group, but to the weighted average of all $k$ groups combined. When $k=2$, $\bar F^w$ is a weighted mixture of $\hat F_1^w$ and $\hat F_0^w$, and each statistic below reduces algebraically to its two-group form (Sections 2.2--2.4).

As before, weights are required to be strictly positive and need not be normalized to any particular scale; every formula below enters weights only as ratios, so any common rescaling of a group's weights leaves every result unchanged.

\subsection{The $k$-Sample Kolmogorov--Smirnov Test}

\subsubsection{Unweighted}

Kiefer \citep{kiefer1959} generalized the two-sample KS statistic to $k$ samples by replacing the single pairwise comparison $|\hat F_1(x)-\hat F_0(x)|$ with a sum, across all $k$ groups, of each group's squared discrepancy from the pooled ECDF:
\begin{equation}
T \;=\; \sup_{x\in\mathbb{R}} \sum_{j=1}^{k} n_j \left[\hat F_j(x) - \bar F(x)\right]^2.
\end{equation}
At $k=2$, this reduces algebraically to a rescaled version of the classical two-sample statistic: $T = \frac{n_1 n_0}{n_1+n_0} D^2$, where $D$ is the two-sample KS statistic defined by \citet{linden2026arxiv}. This scaling factor --- the familiar $n_1n_0/(n_1+n_0)$ term common to two-sample distributional tests --- is a consequence of Kiefer's construction and does not affect any permutation-based inference built on $T$, since it depends only on the (fixed) group sizes and multiplies the observed and every permuted statistic identically.

Because this scaling means $T$ is not on the same numerical scale as $D$ even when $k=2$, and because a $k$-sample statistic being fully backward-compatible with its two-group predecessor is a design goal of this extension (Section 2.6), we define the reported $k$-sample KS statistic as $D$ itself, not the rescaled $T$, whenever exactly two groups are compared, and as $T$ only for $k>2$. This is a choice about \emph{which} equivalent statistic to report, not a second, different statistic: $T$ and $D$ order every possible dataset identically at $k=2$ (both are strictly increasing functions of the same underlying maximal discrepancy), so this choice affects only the reported number, never the resulting inference.

\subsubsection{Weighted}

The weighted $k$-sample KS statistic substitutes weighted ECDFs and weighted group sizes throughout:
\begin{equation}
T^{\,w} \;=\; \sup_{x\in\mathbb{R}} \sum_{j=1}^{k} W_j \left[\hat F_j^{\,w}(x) - \bar F^{\,w}(x)\right]^2,
\end{equation}
where $W_j = \sum_i w_{ji}$ is the sum of weights in group $j$. As in Section 2.2.1, we report the weighted two-sample $D^w$ directly when $k=2$, and $T^w$ only for $k>2$; $T^w$ reduces to $T$ when all weights equal 1, and $D^w$ reduces to $D$ under the same condition, exactly as before.

\subsection{The $k$-Sample Cram\'er--von Mises Test}

\subsubsection{Unweighted}

Kiefer's \citep{kiefer1959} $k$-sample generalization of CVM replaces the pairwise squared discrepancy with the same group-to-pooled sum used for KS, but summed (rather than maximized) across every distinct value in the combined sample:
\begin{equation}
W \;=\; \sum_{x} \sum_{j=1}^{k} n_j \left[\hat F_j(x) - \bar F(x)\right]^2.
\end{equation}
As with $T$, $W$ reduces at $k=2$ to a rescaled version of the two-sample CVM statistic; we again report the unscaled two-sample statistic directly at $k=2$ and $W$ only for $k>2$, for the same backward-compatibility reasons.

\subsubsection{Weighted}

\begin{equation}
W^{\,w} \;=\; \sum_{x} \sum_{j=1}^{k} W_j \left|\hat F_j^{\,w}(x) - \bar F^{\,w}(x)\right|^{p},
\end{equation}
with $p=2$ the conventional choice, generalizing the configurable exponent already used in the two-group case. $W^w$ reduces to $W$, and to the two-sample $CVM^w$ at $k=2$, when all weights equal 1.

\subsection{The $k$-Sample Anderson--Darling Test}

\subsubsection{Unweighted}

Scholz and Stephens \citep{scholz1987} generalized the discrete two-sample AD statistic of Pettitt \citep{pettitt1976} to $k$ samples. For a combined sample of $N=\sum_j n_j$ observations, with $x_{(1)} < x_{(2)} < \cdots < x_{(m)}$ the $m \le N$ distinct values observed and $h_l$ the number of raw (pre-deduplication) observations sharing value $x_{(l)}$,
\begin{equation}
A^{2} \;=\; \frac{N-1}{N} \sum_{l=1}^{m-1} h_l\, \frac{\sum_{j=1}^{k} n_j\left[\hat F_j(x_{(l)}) - \bar F(x_{(l)})\right]^{2}}{N\,\bar F(x_{(l)})\left[1-\bar F(x_{(l)})\right] - h_l/4}.
\end{equation}
The $-h_l/4$ term is Scholz and Stephens's correction for the discrete (tied-data) case; it vanishes as sample sizes grow, recovering the continuous-case statistic. We confirmed algebraically that this expression reduces exactly to Pettitt's \citep{pettitt1976} two-sample statistic at $k=2$.

\subsubsection{Weighted}

As in the two-group extension, three substitutions generalize $A^2$ to the weighted setting: weighted ECDFs and weighted pooled ECDF in place of their unweighted counterparts; the raw sample size $N$ replaced, in the variance-standardization term, by the total Kish \citep{kish1965} effective sample size $n_e = \sum_j n_{e,j}$, $n_{e,j} = W_j^2/\sum_i w_{ji}^2$; and the tie multiplier $h_l$ retained as the raw, unweighted count of observations sharing $x_{(l)}$, for the same reasons given by \citet{linden2026arxiv}. With these substitutions,
\begin{equation}
A^{2,w} \;=\; \frac{n_e-1}{n_e} \sum_{l=1}^{m-1} h_l\, \frac{\sum_{j=1}^{k} W_j\left[\hat F_j^{\,w}(x_{(l)}) - \bar F^{\,w}(x_{(l)})\right]^{2}}{n_e\,\bar F^{\,w}(x_{(l)})\left[1-\bar F^{\,w}(x_{(l)})\right] - h_l/4}.
\end{equation}

\textbf{Backward compatibility at $k=2$.} Unlike the KS and CVM statistics above, $A^{2,w}$ at $k=2$ is \emph{not} simply a rescaled version of the two-group weighted AD statistic reported by \citet{linden2026arxiv}: that two-group statistic, for reasons of computational convenience, is expressed algebraically equivalent to Pettitt's two-sample statistic only up to a constant depending on group sizes, and omits both the tie correction $h_l/4$ and the leading $(n_e-1)/n_e$ factor present in the discrete Scholz--Stephens formulation. To ensure that the $k$-sample statistic is a genuine superset of the two-group statistic --- reproducing not only the same permutation p-value but the identical statistic --- the $k=2$ case is defined using the exact expression from the two-group extension, and the equation above is used only for $k>2$. This is the same design choice described in Section 2.2.1 for KS, applied here as well, and is more consequential for AD than for KS or CVM precisely because the two-group AD statistic and the general Scholz--Stephens formula are not related by a simple constant.

\subsection{Permutation Inference}

All $k$-sample statistics above are compared to a permutation-based, rather than tabulated asymptotic, null distribution, using a direct extension of the procedure described by \citet{linden2026arxiv}. Each of $R$ replicates randomly reshuffles the $N=\sum_j n_j$ group labels among the $N$ observations, preserving the original $k$ group sizes $n_1,\ldots,n_k$, while each observation's value and weight remain fixed to that observation. The statistic is recomputed on each relabeled dataset, and the permutation p-value is
\begin{equation}
p \;=\; \frac{1+\#\{r: T_r^* \ge T_{\text{obs}}\}}{R+1}.
\end{equation}
This is identical in form to the two-group procedure; the only change is that group labels are now drawn from $k$ categories rather than 2, with all $k$ group sizes held fixed across permutations. As before, weight is treated as a fixed attribute of the observation, so the procedure applies without modification to any weighting scheme.

\subsection{Post-Hoc Pairwise Comparisons}

An omnibus rejection of $H_1$ indicates that at least one group's distribution differs from the others, but does not by itself indicate which group or pair of groups. We address this with an optional post-hoc procedure that runs the corresponding \textbf{two-sample} test --- Sections 2.2--2.4's $k=2$ case, identical in form to the original two-group test --- on every one of the $\binom{k}{2}$ pairs of groups, using the same weighting, the same number of permutation replicates, and a single continuous random-number stream shared with the omnibus test, so that one seed reproduces the omnibus result and every pairwise result together.

Because $\binom{k}{2}$ tests inflate the family-wise probability of at least one false positive, raw pairwise p-values are additionally adjusted using a choice of four standard methods: Bonferroni and \v{S}id\'ak \citep{sidak1967}, both simple single-step corrections controlling the family-wise error rate; Holm's \citep{holm1979} step-down procedure, uniformly more powerful than Bonferroni while retaining the same family-wise guarantee; and the Benjamini--Hochberg \citep{benjamini1995} procedure, which controls the false discovery rate rather than the family-wise error rate and is typically the least conservative of the four. All four are reported simultaneously rather than requiring the user to select one in advance.

Because each pairwise test is a $k=2$ case, it uses the exact two-group statistic of Sections 2.2--2.4 (not the $k$-sample $T^w$/$W^w$/$A^{2,w}$ restricted to two groups), so a post-hoc pairwise result is numerically identical to the two-group statistic computed on that pair directly.

Each individual test call, omnibus or pairwise, costs $O(RN\log N)$: an $O(N\log N)$ sort of the pooled sample, repeated across $R$ permutation replicates. The post-hoc procedure adds $\binom{k}{2}$ such calls; bounding each conservatively at the same order (each pairwise call in fact sorts a smaller subsample than the full pooled data) gives an overall post-hoc cost of $O\!\left(k^2 RN\log N\right)$, growing quadratically in the number of groups. For the small numbers of groups typical of covariate-balance applications, this remains practical; for substantially larger $k$, or when the post-hoc procedure is applied across many covariates within the same analysis or repeated inside a larger computational pipeline, the quadratic scaling is worth accounting for when choosing $R$.

\subsection{Simulation Study Design}

The two-group simulation study of \citet{linden2026arxiv} used a common data-generating process (DGP) across four scenarios --- Type~I error, and power against centrally located, tail-located, and diffuse discrepancies --- to test whether each test's known theoretical sensitivity persisted under weighting. We extend that design to $k=3$ groups in a way that preserves direct comparability with the original study while introducing the minimum structural change needed to have a genuine $k>2$ comparison.

\subsubsection{Group Structure}

Rather than making all three groups mutually different, which would confound ``is there a discrepancy'' with ``where between which groups is it,'' we designate one group as \textbf{discrepant} and the other two as \textbf{reference groups}, both drawn from the identical distribution. This isolates the same question the two-group study asked --- does a test detect a discrepancy of a given type and location? --- while adding a genuine three-group comparison and, as a byproduct, a natural check on whether the two reference groups are correctly \emph{not} distinguished from each other.

\subsubsection{Allocation}

The original study allocated $n_1/n=0.4$ to the (discrepant) treatment group and $n_0/n=0.6$ to the (reference) comparison group. We preserve the discrepant group's share exactly ($n_3/n=0.4$) and split the reference share evenly across the two reference groups ($n_1/n=n_2/n=0.3$ each), keeping the discrepant-vs-reference comparisons as close as possible in scale to the original two-group power curves.

\subsubsection{Data-Generating Process}

For a total sample size $n$, each observation is assigned a quantile position $p_{ji}\sim U(0,1)$ and an outcome
\begin{equation}
y_{ji} \;=\; \mu + \sigma\,\Phi^{-1}(p_{ji}) + \mathbb{1}(j=3)\,\delta(p_{ji}),
\end{equation}
with $\mu=50$, $\sigma=10$, and $\delta(\cdot)$ the same scenario-specific shift function as the original study (Table~1), applied only to the discrepant group ($j=3$) --- a direct extension of the original DGP, with the single treatment indicator $\mathbb{1}(g=1)$ replaced by $\mathbb{1}(j=3)$.

\subsubsection{Case Weights}

Assigned independently of $y$, $j$, and $p$, using the identical lognormal construction as the original study,
\begin{equation}
w_{ji} \;=\; \exp(\eta_{ji}), \qquad \eta_{ji} \sim N\!\left(-\tfrac{1}{2}\sigma_\eta^2,\ \sigma_\eta^2\right), \qquad \sigma_\eta^2 = \ln\!\left(1+c^2\right),\ \ c^2 = \frac{1}{r_e}-1,
\end{equation}
targeting a Kish effective-sample-size ratio $r_e = n_e/n$. The main study fixed $r_e=0.6$ throughout; a follow-up sensitivity check, addressing the same question as the original study's weight-variability sensitivity check, re-examined all four scenarios at a single representative sample size ($n=2000$, with $n_1=n_2=600$, $n_3=800$) across $r_e \in \{0.2, 0.4, 0.8, 0.9\}$.

\subsubsection{Common Design Elements}

Total sample size $n\in\{1000,2000,3000,4000\}$, yielding 16 scenario-by-sample-size combinations at $r_e=0.6$, each replicated 2000 times; every test's permutation p-value used $R=1000$ label-permutation replicates. The weight-variability sensitivity check added 4 further $r_e$ values at the single sample size $n=2000$, again with 2000 replications per condition and $R=1000$ permutation replicates per test. All analyses were conducted in Stata (version 19) using the community-contributed Stata commands \texttt{kstest} \citep{lindenkstest2026}, \texttt{adtest} \citep{lindenadtest2025}, and \texttt{cvmtest} \citep{lindencvmtest2025}. No post-hoc pairwise comparisons were performed in this simulation study; it evaluates only the omnibus tests' Type~I error and power, extending the original study's own scope rather than additionally validating the post-hoc procedure of Section 2.6, which is a separate methodological question left to future work (see Discussion).

\subsubsection{Paired Comparisons}

As in the original study, because all three tests are computed on the identical simulated dataset within each replicate, their rejection outcomes are correlated. For each pair of tests and each scenario-by-sample-size combination, the per-replicate difference in rejection outcome at $\alpha=0.05$ was computed directly, and a paired $z$-test (mean difference divided by its standard error, incorporating the within-replicate correlation) assessed whether one test significantly outperformed another.

\section{Results}

Full results for the $k=3$ extension are reported in Sections 3.1--3.6 below. Every power scenario shows systematically lower power at $k=3$ than the corresponding $k=2$ result reported by \citet{linden2026arxiv}, for every test; the mechanism behind this pattern is developed in the Discussion. Type~I error, by contrast, is essentially unchanged between the two studies.

\subsection{Type I Error}

Table~2 reports empirical rejection rates at $\alpha=0.05$ under the null across the four sample sizes, at $k=3$. With 2000 replications per cell, all three tests clustered tightly around the nominal $5\%$ level at every sample size, essentially indistinguishable from the two-group study's own Type~I error results both in magnitude and in the absence of any systematic trend across $n$. None of the corresponding paired-difference tests reached significance except CVM--KS at $n=2000$ ($p=.061$), which does not survive even a nominal threshold and is consistent with chance alone across the 12 null-condition comparisons tested.

\subsection{Power Against a Centrally Located Discrepancy}

Table~3a reports empirical power against the centrally located discrepancy. KS was the most powerful test at every sample size, followed by AD, with CVM trailing both --- the identical ordering documented for two groups. Paired-difference tests (Table~4) confirmed all three pairwise orderings significant at every sample size (all $p<.001$), including AD$>$CVM at $n=1000$, where the two-group study had required $n=2000$ for that specific comparison to reach significance.

\subsection{Power Against a Tail-Located Discrepancy}

Table~3b reports empirical power against the tail-located discrepancy. AD was overwhelmingly the most powerful test, rising from near its own null rejection rate at $n=1000$ to $.618$ at $n=4000$, while KS and CVM both remained close to their own Type~I error rates throughout --- mirroring the two-group study's tail-scenario finding without qualification. AD significantly exceeded both alternatives at every sample size (all $p<.001$). CVM held a small, inconsistent edge over KS (significant at $n=1000,2000$; not at $n=3000,4000$), the same small, second-order effect documented previously, though its direction here (CVM slightly ahead of KS, where the earlier study found the reverse at its largest sample size) is not itself a substantive finding given its inconsistency across $n$.

\subsection{Power Against a Diffuse Discrepancy}

Table~3c reports empirical power against the diffuse discrepancy. Both AD and CVM clearly outperformed KS at every sample size $n \ge 2000$ (both comparisons significant, $p<.001$), with all three tests statistically indistinguishable at $n=1000$. AD held a small but consistently significant edge over CVM from $n=2000$ onward ($p=.010$, $.008$, $<.001$ at $n=2000,3000,4000$) --- the same ``AD and CVM comparable, with a modest AD advantage from moderate sample size onward'' pattern as the two-group study.

\subsection{Summary of Paired Comparisons}

Table~4 reports the full set of paired differences underlying Sections 3.2--3.4. Of the 12 Type-I-error paired comparisons, none reached significance, consistent with chance alone and providing no evidence of a systematic size difference among the three tests at $k=3$.

\subsection{Weight-Variability Sensitivity}

Table~5 reports empirical Type~I error and power at the fixed sample size $n=2000$, across five levels of weight variability, $r_e \in \{0.2,0.4,0.6,0.8,0.9\}$. The comparative ordering documented in Sections 3.2--3.4 held at every $r_e$ level tested, for all three scenarios, without exception: KS $>$ AD $>$ CVM throughout the central scenario, AD dominant throughout the tail scenario with KS and CVM both remaining near their own null rate, and AD holding a small, consistent edge over CVM throughout the diffuse scenario.

\textbf{One notable departure from the two-group study.} \citet{linden2026arxiv} found Type~I error inflated to roughly 30\% above nominal at the most severe weight variability examined ($r_e=0.2$: $.064$--$.067$ across all three tests, versus a nominal $.05$). At $k=3$, Type~I error at $r_e=0.2$ was $.044$--$.048$ across all three tests --- if anything marginally below nominal, and well within the range observed at every other $r_e$ level. With 2000 replications per cell (standard error $\approx.005$ at $p\approx.05$), this is not attributable to reduced precision relative to the original study. Whether this reflects a genuine difference between the two-group and $k$-sample permutation procedures under extreme weight variability, or is specific to the particular discrepant-group DGP used here, is addressed further in the Discussion.

\section{Applied Example}

\subsection{Study Context and Data}

We illustrate the $k$-sample tests using data from a disease management (DM) program for patients with congestive heart failure examined in two prior methodological papers introducing and comparing multivalued-treatment adjustment approaches on this dataset \citep{linden2014,linden2016multivalued}. The program was implemented in a large health plan in the western United States; individuals with the condition were invited to enroll and, on agreeing to participate, received one of two interventions based on the program nurse's subjective assessment of patient needs and preferences: periodic telephone calls from a nurse to discuss self-management behaviors, or remote tele-monitoring (RTM) involving daily electronic transmission of the participant's disease-related symptoms with nurse follow-up when symptoms suggested an impending exacerbation \citep{lindenblackbox2006}.

Health plan members with the condition who did not enroll served as a non-participant comparison group. The retrospectively collected data consist of 6612 non-participants (Control), 654 participants in the telephonic intervention (Calls), and 705 participants in the RTM intervention, each with 12 months of pre-intervention data. We use these data solely to illustrate the statistical methods for assessing covariate balance, and not to assess the program's effectiveness.

\subsection{Covariate Balance Assessment}

Following \citet{linden2014} and \citet{linden2016multivalued}, marginal mean weights through stratification (MMWS) were constructed using the generalized approach for nominal treatments described in \citet{linden2014}: a generalized propensity score for each of the three treatment levels was estimated by multinomial logistic regression. Rather than re-examining those same covariates, for illustration, we estimate the propensity score and apply the $k$-sample distributional tests developed in this paper to a covariate that played no role in constructing the original weights: a medical risk score reflecting each patient's model-predicted near-term medical costs, based on prior health care utilization history.

\subsection{Results}

Before weighting, all three tests found overwhelming evidence of distributional imbalance in the medical risk score across the three groups: the omnibus KS, AD, and CVM tests were all significant at $p=.001$ (the floor imposed by 1000 permutation replicates), as was every one of the nine pairwise comparisons (three pairs $\times$ three tests), before and after adjustment for multiple comparisons. This confirms that, absent adjustment, the three groups differ substantially on the distribution of risk scores. Figure~1 displays the unweighted empirical CDFs of the medical risk score for the three groups, with the location of the omnibus KS statistic marked directly on the plot; the RTM group's curve visibly diverges from the other two across much of the score's distribution.

After MMWS weighting, this picture reversed. Table~6 reports the omnibus test results for both the unweighted and MMWS-weighted concurrent medical risk score; none of the three weighted omnibus tests reached significance at $\alpha=.05$, a marked contrast to the unweighted results in the same table. Figure~2 displays the corresponding MMWS-weighted empirical CDFs: the three groups' curves are visibly nearly coincident throughout the distribution, consistent with the near-null weighted KS statistic. Table~7 reports the corresponding weighted post-hoc pairwise comparisons: none of the nine pairwise comparisons (three pairs $\times$ three tests) reached significance either, before or after adjustment for multiple comparisons (the unweighted pairwise comparisons are not tabulated separately, since all nine were significant at $p=.001$, the same floor value reported for the unweighted omnibus tests). The weighting procedure appears to have balanced a clinically meaningful summary measure of near-term medical cost risk.

\section{Discussion}

\subsection{Principal Findings}

This study extended the Kolmogorov--Smirnov, Cram\'er--von Mises, and Anderson--Darling tests to an arbitrary number of weighted groups $k \ge 2$, introduced a post-hoc pairwise comparison procedure with a choice of multiple-comparison adjustments, and evaluated the $k$-sample tests at $k=3$, plus an applied example. Four findings stand out. First, all three tests controlled Type~I error close to nominal across sample sizes, matching the two-group study's own results. Second, all three comparative advantages established for two groups --- KS for a centrally located discrepancy, AD for a tail-located discrepancy, AD and CVM comparably for a diffuse discrepancy --- were confirmed at $k=3$, if anything more cleanly than at $k=2$ for the diffuse scenario. Third, every power scenario showed systematically lower power at $k=3$ than at the matching $k=2$ condition, for every test, while Type~I error was essentially unchanged; Section~5.2 explains why. Fourth, the applied example showed the post-hoc procedure resolving unambiguous baseline imbalance (every test significant at the permutation floor) into a result with no detectable residual imbalance after weighting.

\subsection{Mechanistic Interpretation}

The centrally located and tail-located results follow from the same mechanisms established at $k=2$ \citep{linden2026arxiv}: KS's dependence on a single largest gap, and AD's variance term being smallest at the center and largest in the tails. Extending to $k$ weighted groups did not disturb either mechanism.

This power reduction has a more specific source, and it is not the one intuition first suggests. Pooling the discrepant group into the reference curve $\bar F^w$ might seem to dilute the measured discrepancy, but it does not: for one discrepant group of share $\pi$ against $k-1$ identical reference groups, Kiefer's $T$ reduces algebraically to $n\pi(1-\pi)(F_3-F_{\text{ref}})^2$, exactly the value a two-group comparison with the same $\pi$ and $n$ would give --- the population-level signal is unchanged. The reduction instead comes from the null distribution: Kiefer's and Scholz-Stephens's asymptotic theory index the null distributions of all three statistics by $k-1$ degrees of freedom, and these families are stochastically increasing in that index, so the critical value needed to reject grows with $k$. This is a permutation-based analogue of an omnibus ANOVA $F$-test losing power for a single deviant group as more (non-differing) groups are added --- a structural property of the omnibus construction, not an artifact of weighting or of the discrepancy types examined here.

\subsection{Practical Advice for Investigators}

The test-choice guidance from the two-group setting carries over unchanged: KS when a centrally concentrated imbalance is suspected, AD when it may be in the tails, and AD as a reasonable general-purpose default otherwise, given its consistent edge over CVM and its outsized tail sensitivity.

The mechanism in Section~5.2 adds a consideration with no two-group analogue: because every omnibus test loses power for a single deviant group as more groups are added, an investigator who specifically suspects one group among several should not expect the omnibus test to be as sensitive as a direct two-group comparison. The post-hoc procedure of Section~2.6 provides that direct comparison without leaving the omnibus framework, since each pairwise statistic carries none of the omnibus test's power penalty; routine reporting of post-hoc results, not only when the omnibus test rejects, is often informative whenever more than two groups are compared, particularly when substantive interest focuses on identifying which groups differ.

Left unadjusted, the $\binom{k}{2}$ raw pairwise p-values inflate the chance of at least one false positive well above the nominal level: at $k=3$, three comparisons raise the probability of at least one spurious rejection from $5\%$ to approximately $14\%$ under independence ($1-(1-.05)^3$), and this inflation grows quickly with $k$. Among the four adjustments offered (Section~2.6), Bonferroni and \v{S}id\'ak are the simplest and most conservative; Holm's step-down procedure controls the same family-wise error rate with uniformly more power and is a reasonable default when any single false positive among the pairwise comparisons would be consequential, for instance when confirming that no pair among several treatment arms shows residual imbalance. The Benjamini--Hochberg procedure controls the false discovery rate rather than the family-wise error rate and is better suited to a more exploratory use of the post-hoc procedure, such as screening many covariates or many groups at once, where an investigator plans to follow up individually on whichever pairs are flagged and can tolerate a controlled proportion of false leads among them in exchange for greater power. The post-hoc procedure's computational cost, discussed in Section~2.6, is worth weighing against this choice as $k$ grows.

Because the omnibus and post-hoc statistics are standardized against different reference distributions --- the omnibus tests compare each group to the pooled distribution of all $k$ groups, while each pairwise test compares only the two groups involved --- the two need not agree, and disagreement between them is not itself evidence that either result is wrong. An omnibus rejection with no individually significant pair can arise when a real discrepancy is spread thinly enough across several groups that no single pairwise comparison captures it in isolation. A non-significant omnibus result alongside a significant pairwise comparison can arise from the power penalty of Section~5.2, particularly when only one pair among several groups is truly imbalanced and the remaining groups dilute the omnibus test's sensitivity to it without eliminating the discrepancy at the pairwise level. Investigators encountering such disagreement should treat the pairwise results, not the omnibus result, as the more direct evidence about which specific groups differ, and should not expect the omnibus test alone to substitute for post-hoc comparisons whenever the applied question concerns particular groups rather than overall homogeneity.

As in the two-group setting, AD's tie multiplier is retained as a raw, unweighted count, so a large tie block in a semi-continuous covariate can drive a significant AD result on its own; investigators should inspect such covariates directly rather than relying on the test statistic alone, and should report the process used to select a test \citep{lindenroberts2005,linden2003dmaa}.

\subsection{Limitations}

Four limitations are worth noting. First, as in the two-group study, the data-generating process used a normal base distribution and a single fixed effect size per scenario; whether the findings extend to non-normal distributions or other effect sizes was not assessed. Second, the $k=3$ design examined only one discrepancy structure (one discrepant group against $k-1$ identical references); other structures possible only at $k>2$ remain untested, and whether the power reduction documented in Sections~3.2--3.4 scales linearly, sublinearly, or otherwise with increasing $k$ remains unknown. Third, the post-hoc procedure's own Type~I error and power were not evaluated by simulation here, only illustrated in the applied example, whose before/after pattern --- overwhelming imbalance resolving into a uniformly null result --- does not showcase the more diagnostically interesting case of an omnibus rejection with only a subset of pairs remaining significant after adjustment; a dataset exhibiting that pattern would better illustrate the post-hoc procedure's localization value, and the procedure's own operating characteristics remain a natural target for follow-up simulation. Fourth, the post-hoc procedure's computational cost (Section~2.6) grows quadratically in $k$; this remains modest for the group counts typical of covariate-balance applications but is worth accounting for at larger $k$ or when applied across many covariates.

\subsection{Conclusion}

This paper extends the Kolmogorov--Smirnov, Cram\'er--von Mises, and Anderson--Darling tests to an arbitrary number of weighted groups, with each $k$-sample statistic reducing exactly to its two-group counterpart at $k=2$, and introduces a post-hoc pairwise procedure for localizing which groups differ once an omnibus test rejects. Both were supported by a $k=3$ simulation study: Type~I error remained close to nominal, and all three comparative advantages established for two groups were confirmed intact. A new finding specific to this extension is that every omnibus test loses power, for a fixed single-group discrepancy, as non-differing groups are added to the comparison --- a consequence of additional degrees of freedom in the null distribution rather than diluted signal --- making the post-hoc procedure, whose pairwise statistics carry no such penalty, the more sensitive tool whenever a specific group's imbalance is the concern. The applied example illustrated this directly, with MMWS weighting resolving unambiguous imbalance into a well balanced covariate. AD remains a reasonable general-purpose default with more than two groups, as it was with two; a natural direction for future work is extending the same framework to other $k$-sample distributional tests, such as the Kuiper \citep{Kuiper1960} and Wasserstein \citep{Ramdas2017} statistics, and characterizing how the power reduction and post-hoc performance documented here scale beyond $k=3$.

\bibliographystyle{apalike}
\bibliography{refs}

\clearpage

\begin{table}[htbp]
\centering
\caption{Simulation Study Design ($k=3$)}
\small
\begin{tabular}{@{}p{0.28\textwidth}p{0.60\textwidth}@{}}
\toprule
Purpose & Shift function $\delta(p)$, applied to discrepant group ($j=3$) \\
\midrule
Type I error                & $\delta \equiv 0$ \\
Power: centrally located    & Localized: $-\Delta$ at $p_L=.25$, $+\Delta$ at $p_U=.75$, $w=.05$, $\Delta=6$ \\
Power: tail-located         & Localized: $-\Delta$ at $p_L=.05$, $+\Delta$ at $p_U=.95$, $w=.02$, $\Delta=50$ \\
Power: diffuse              & $\delta \equiv 1$ (uniform shift) \\
\bottomrule
\end{tabular}

\vspace{4pt}
\parbox{0.9\textwidth}{\footnotesize Note: all four scenarios were run at total sample size $n \in \{1000, 2000, 3000, 4000\}$, with $n_1/n=n_2/n=0.3$ (reference groups), $n_3/n=0.4$ (discrepant group) throughout. Case weights follow a lognormal distribution targeting a Kish effective-sample-size ratio of $r_e=0.6$ (see text), assigned independently of the DGP. Total scenario-by-sample-size combinations: $4 \times 4 = 16$; 2000 replications per combination; $R=1000$ permutation replicates per test per replicate.}
\end{table}

\clearpage

\begin{table}[htbp]
\centering
\caption{Empirical Type I Error Rates ($\alpha=0.05$), $k=3$}
\begin{tabular}{@{}lccc@{}}
\toprule
$n$ & KS & AD & CVM \\
\midrule
1000 & .048 (.039, .057) & .051 (.041, .060) & .051 (.041, .060) \\
2000 & .051 (.041, .061) & .049 (.039, .058) & .044 (.035, .053) \\
3000 & .057 (.047, .067) & .052 (.043, .062) & .056 (.046, .066) \\
4000 & .050 (.040, .060) & .047 (.038, .057) & .047 (.037, .056) \\
\bottomrule
\end{tabular}

\vspace{4pt}
\parbox{0.85\textwidth}{\footnotesize Note: 2000 replications per cell; values in parentheses are 95\% confidence intervals ($\hat p \pm 1.96\sqrt{\hat p(1-\hat p)/2000}$).}
\end{table}

\clearpage

\begin{table}[htbp]
\centering
\caption{Empirical Power Against Three Types of Distributional Discrepancy, $k=3$}
\footnotesize
\begin{tabular}{@{}llccc@{}}
\toprule
Scenario & $n$ & KS & AD & CVM \\
\midrule
Centrally located (Table 3a) & 1000 & .255 (.235, .274) & .113 (.099, .127) & .094 (.081, .107) \\
                  & 2000 & .674 (.653, .695) & .271 (.252, .291) & .207 (.190, .225) \\
                  & 3000 & .918 (.905, .930) & .506 (.484, .527) & .391 (.369, .412) \\
                  & 4000 & .984 (.979, .989) & .714 (.694, .734) & .579 (.557, .601) \\
\addlinespace
Tail-located (Table 3b) & 1000 & .045 (.036, .055) & .090 (.077, .102) & .055 (.045, .065) \\
             & 2000 & .050 (.040, .060) & .181 (.164, .197) & .060 (.050, .070) \\
             & 3000 & .054 (.045, .064) & .352 (.332, .373) & .063 (.052, .074) \\
             & 4000 & .064 (.053, .074) & .618 (.597, .639) & .069 (.058, .081) \\
\addlinespace
Diffuse (Table 3c) & 1000 & .138 (.122, .153) & .156 (.141, .172) & .156 (.141, .172) \\
        & 2000 & .226 (.208, .244) & .281 (.261, .301) & .269 (.250, .288) \\
        & 3000 & .330 (.309, .351) & .389 (.368, .410) & .376 (.355, .397) \\
        & 4000 & .444 (.422, .466) & .531 (.509, .553) & .510 (.488, .532) \\
\bottomrule
\end{tabular}

\vspace{4pt}
\parbox{0.9\textwidth}{\footnotesize Note: 2000 replications per cell; rejection rate at $\alpha=0.05$ using each test's permutation p-value ($R=1000$ replicates). Values in parentheses are 95\% confidence intervals. See Table~4 for paired-difference significance tests. Sub-panels a/b/c correspond to the three discrepancy scenarios and may be split into separate table numbers at final formatting.}
\end{table}

\clearpage

\begin{table}[htbp]
\centering
\caption{Paired Differences in Rejection Rate Between Tests, $k=3$}
\footnotesize
\begin{tabular}{@{}llrrrrrr@{}}
\toprule
 & & \multicolumn{2}{c}{AD $-$ KS} & \multicolumn{2}{c}{CVM $-$ KS} & \multicolumn{2}{c}{AD $-$ CVM} \\
\cmidrule(lr){3-4}\cmidrule(lr){5-6}\cmidrule(lr){7-8}
Scenario & $n$ & $\Delta$ & $p$ & $\Delta$ & $p$ & $\Delta$ & $p$ \\
\midrule
Type I error & 1000 & .003 & .541 & .003 & .515 & .000 & 1.000 \\
             & 2000 & $-$.003 & .541 & $-$.007 & .061 & .005 & .095 \\
             & 3000 & $-$.005 & .323 & $-$.001 & .808 & $-$.004 & .262 \\
             & 4000 & $-$.003 & .475 & $-$.004 & .262 & .001 & .655 \\
\addlinespace
Central & 1000 & $-$.142 & $<$.001 & $-$.161 & $<$.001 & .019 & $<$.001 \\
        & 2000 & $-$.403 & $<$.001 & $-$.467 & $<$.001 & .064 & $<$.001 \\
        & 3000 & $-$.412 & $<$.001 & $-$.527 & $<$.001 & .115 & $<$.001 \\
        & 4000 & $-$.270 & $<$.001 & $-$.405 & $<$.001 & .135 & $<$.001 \\
\addlinespace
Tail & 1000 & .044 & $<$.001 & .010 & .022 & .035 & $<$.001 \\
     & 2000 & .131 & $<$.001 & .010 & .015 & .121 & $<$.001 \\
     & 3000 & .298 & $<$.001 & .009 & .059 & .290 & $<$.001 \\
     & 4000 & .555 & $<$.001 & .006 & .257 & .549 & $<$.001 \\
\addlinespace
Diffuse & 1000 & .019 & .002 & .019 & .001 & .000 & 1.000 \\
        & 2000 & .055 & $<$.001 & .043 & $<$.001 & .012 & .010 \\
        & 3000 & .059 & $<$.001 & .046 & $<$.001 & .013 & .008 \\
        & 4000 & .087 & $<$.001 & .066 & $<$.001 & .021 & $<$.001 \\
\bottomrule
\end{tabular}

\vspace{4pt}
\parbox{0.95\textwidth}{\footnotesize Note: $\Delta$ is the mean per-replicate difference in rejection outcome; $p$-values are from a paired $z$-test accounting for the correlation between tests run on identical simulated data. $p<.001$ shown where the exact value rounds to $.000$ at three decimal places.}
\end{table}

\clearpage

\begin{table}[htbp]
\centering
\caption{Weight-Variability Sensitivity: Empirical Type I Error and Power by Condition and $r_e$ ($n=2000$), $k=3$}
\begin{tabular}{@{}llccc@{}}
\toprule
Condition & $r_e$ & KS & AD & CVM \\
\midrule
Type I error & 0.2 & .046 & .048 & .044 \\
             & 0.4 & .047 & .047 & .045 \\
             & 0.6 & .051 & .049 & .044 \\
             & 0.8 & .042 & .044 & .049 \\
             & 0.9 & .050 & .044 & .046 \\
\addlinespace
Power: centrally located & 0.2 & .189 & .104 & .085 \\
                         & 0.4 & .402 & .164 & .118 \\
                         & 0.6 & .674 & .271 & .207 \\
                         & 0.8 & .875 & .413 & .312 \\
                         & 0.9 & .923 & .504 & .373 \\
\addlinespace
Power: tail-located & 0.2 & .050 & .088 & .058 \\
                    & 0.4 & .046 & .125 & .057 \\
                    & 0.6 & .050 & .181 & .060 \\
                    & 0.8 & .052 & .285 & .058 \\
                    & 0.9 & .049 & .351 & .057 \\
\addlinespace
Power: diffuse & 0.2 & .123 & .148 & .142 \\
              & 0.4 & .174 & .206 & .204 \\
              & 0.6 & .226 & .281 & .269 \\
              & 0.8 & .290 & .369 & .351 \\
              & 0.9 & .325 & .405 & .400 \\
\bottomrule
\end{tabular}

\vspace{4pt}
\parbox{0.9\textwidth}{\footnotesize Note: 2000 replications per cell; sample size fixed at $n=2000$ ($n_1=n_2=600$, $n_3=800$) throughout. The $r_e=0.6$ row reproduces the corresponding $n=2000$ cell from Tables 2--3.}
\end{table}

\clearpage

\begin{table}[htbp]
\centering
\caption{Omnibus Test Results for the Concurrent Medical Risk Score, Unweighted and MMWS-Weighted}
\begin{tabular}{@{}lcccc@{}}
\toprule
 & \multicolumn{2}{c}{Unweighted} & \multicolumn{2}{c}{MMWS-weighted} \\
\cmidrule(lr){2-3}\cmidrule(lr){4-5}
Test & Statistic & $p$ & Statistic & $p$ \\
\midrule
KS (T)   & 21.1973     & .001 & 1.9731    & .1359 \\
AD ($A^2$) & 41.6552   & .001 & 3.0435    & .1499 \\
CVM (W)  & 48211.1229  & .001 & 1028.5214 & .8002 \\
\bottomrule
\end{tabular}

\vspace{4pt}
\parbox{0.85\textwidth}{\footnotesize Note: all tests based on 1000 permutation replicates. $p=.001$ is the floor imposed by 1000 replicates ($1/1001$). Statistic magnitudes are not comparable across tests, or between unweighted and weighted versions of the same test; significance is determined entirely by each test's own permutation null distribution, not by the absolute size of its statistic.}
\end{table}

\clearpage

\begin{table}[htbp]
\centering
\caption{Post-Hoc Pairwise Comparisons for the Concurrent Medical Risk Score, MMWS-Weighted}
\footnotesize
\begin{tabular}{@{}lcccccc@{}}
\toprule
 & \multicolumn{2}{c}{KS} & \multicolumn{2}{c}{AD} & \multicolumn{2}{c}{CVM} \\
\cmidrule(lr){2-3}\cmidrule(lr){4-5}\cmidrule(lr){6-7}
Groups & Stat & $p$ & Stat & $p$ & Stat & $p$ \\
\midrule
Control vs.\ Calls & 0.0489 & .1149 & 60836.7670 & .2378 & 0.9700 & .5524 \\
Control vs.\ RTM   & 0.0386 & .2957 & 70342.4220 & .1449 & 0.6915 & .7013 \\
Calls vs.\ RTM     & 0.0319 & .8811 & 777.4282   & .9500 & 0.1330 & .9670 \\
\bottomrule
\end{tabular}

\vspace{4pt}
\parbox{0.95\textwidth}{\footnotesize Note: raw p-values shown; based on 1000 permutation replicates per pair. None of the nine comparisons reached significance at $\alpha=.05$, so Bonferroni-, \v{S}id\'ak-, Holm-, and Benjamini--Hochberg-adjusted p-values (all higher than the corresponding raw value) are omitted here for space. Corresponding unweighted pairwise comparisons are not tabulated: all nine were significant at $p=.001$, the same floor value reported for the unweighted omnibus tests in Table~6.}
\end{table}

\clearpage

\begin{figure}[htbp]
\centering
\includegraphics[width=0.95\textwidth]{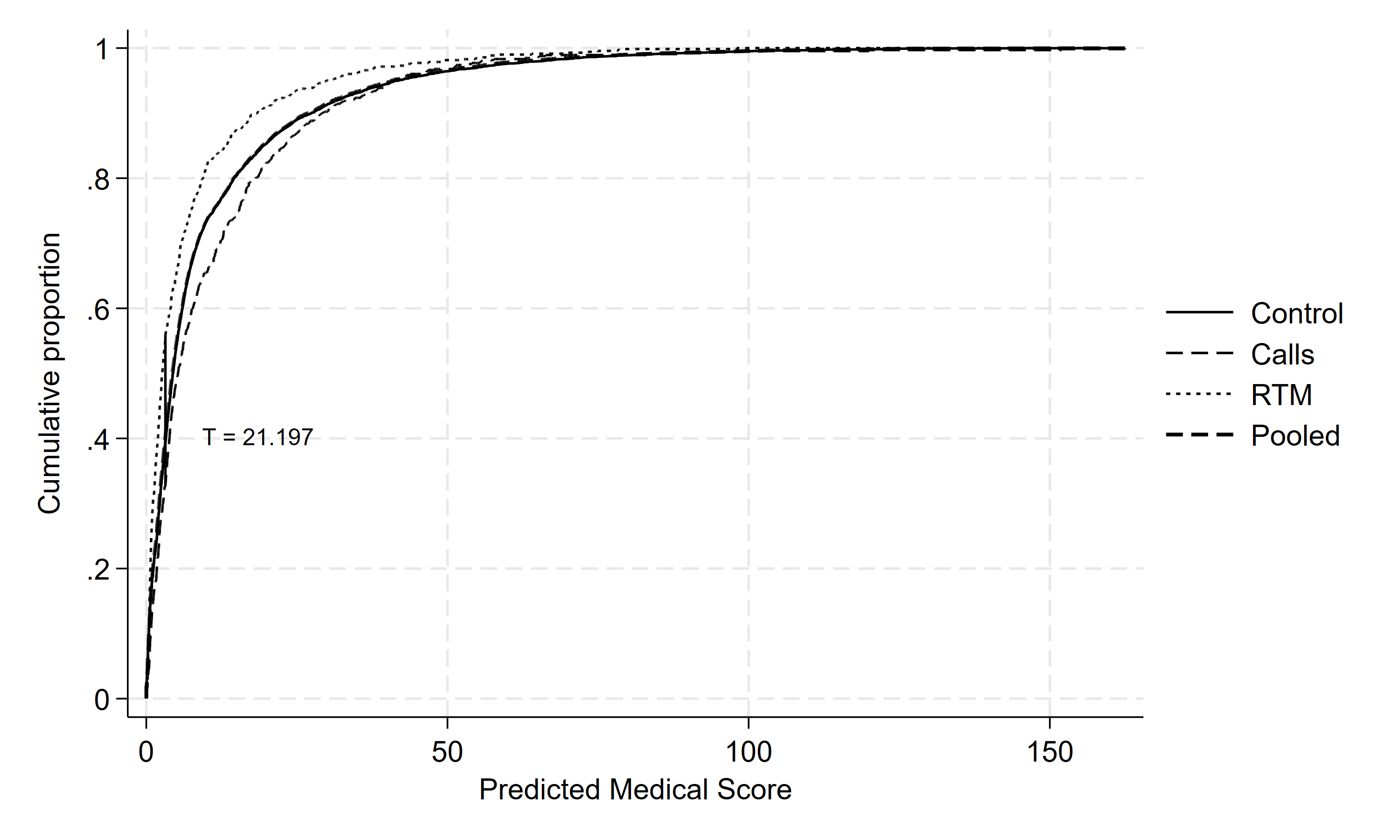}
\caption{Unweighted empirical CDFs of the concurrent medical risk score, Control vs.\ Calls vs.\ RTM. The Kolmogorov--Smirnov $T$ statistic (21.197) is marked at its location.}
\end{figure}

\clearpage

\begin{figure}[htbp]
\centering
\includegraphics[width=0.95\textwidth]{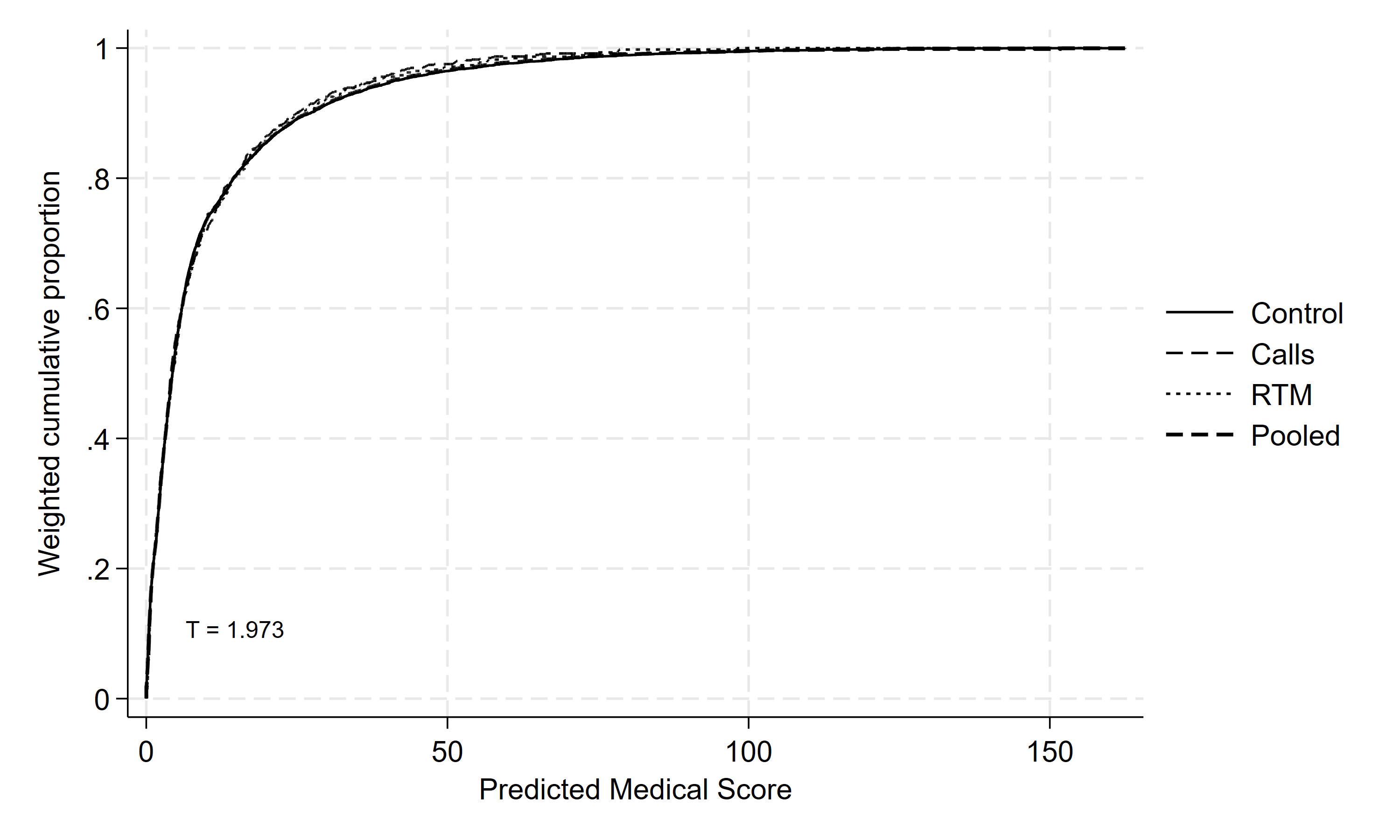}
\caption{MMWS-weighted empirical CDFs of the concurrent medical risk score, Control vs.\ Calls vs.\ RTM. The Kolmogorov--Smirnov $T$ statistic (1.973) is marked at its location.}
\end{figure}

\end{document}